\documentclass[onecolumn,showpacs]{revtex4}

\usepackage{graphicx}
\usepackage{dcolumn}
\usepackage{bm}

\input epsf

\begin{document}

\title{\Large Emergent Universe in Chameleon, $f(R)$ and $f(T)$ Gravity Theories}

\author{\bf  Surajit
Chattopadhyay$^1$\footnote{surajit$_{_{-}}2008$@yahoo.co.in} and
Ujjal Debnath$^2$\footnote{ujjaldebnath@yahoo.com ,
ujjal@iucaa.ernet.in}}

\affiliation{$^1$Department of Computer Application, Pailan
College of Management and Technology, Bengal Pailan Park,
Kolkata-700 104, India.\\
$^2$Department of Mathematics, Bengal Engineering and Science
University,\\ Shibpur, Howrah-711 103, India. }

\date{\today}

\begin{abstract}
In this work, we have considered emergent universe in generalized
gravity theories like Chameleon, $f(R)$ and $f(T)$ gravities. We
have reconstructed the potential of Chameleon field under emergent
scenario of the universe and observed its increasing nature with
evolution of the universe. We have revealed that in the emergent
universe scenario, the equation of state parameter behaves like
quintessence in the case of $f(R)$ gravity and like phantom in the
case of $f(T)$ gravity.
\end{abstract}


\maketitle

\section{\bf\large{Introduction}}

The central paradigm for structure formation in the universe is
the inflationary scenario. Under very general conditions,
inflation is future eternal, in the sense that once inflation has
started, most of the volume of the universe will remain in an
inflating state [1]. We consider the total density
$\Omega(t)=1+\frac{k}{a^{2}H^{2}}$, where, $k$, $a$ and $H$ denote
the curvature, scale factor and Hubble parameter respectively.
Inflation drives the curvature term towards $0$, which does not
imply that $k=0$. However, the standard inflationary model is
based on a flat $(k=0\Longleftrightarrow \Omega_{0}=1)$
Friedmann-Robertson-Walker (FRW) geometry, motivated by the fact
that inflationary expansion rapidly wipes out any original spatial
curvature. Leading alternatives to inflation that are motivated by
recent advances in string/M-theory are the pre-big bang [2] and
ekpyrotic/cyclic [3] scenarios, respectively. The spatial
curvature of the real universe is in principle determined by
observations. The recent WMAP data seems to point to a universe
that is close to (but not quite) flat, with a total density
parameter of $\Omega_{tot}=1.02 \pm 0.02$. Theory will have to
give way to data if the data clearly tell us that $\Omega_{0}>1$
([4],[5]). If $\Omega_{0}$ is taken as 1.02, then the power
spectra of CMB anisotropies and matter can show testable
differences from the standard flat model
[6]. \\

The singularity theorems assume that either the universe has open
space sections, implying $k = 0$ or $-1$; or the Hubble expansion
rate $H =\frac{\dot{a}}{a}$ is bounded away from zero in the past.
There are inflationary universes that evade these constraints and
hence avoid the conclusions of the theorems [7]. Ellis and
Maartens [4] considered closed models in which $k =+1$ and $H$ can
become zero, so that both of the assumptions of the inflationary
singularity theorems are violated. In this paper [4], it was shown
that if $k = +1$ then there are closed inflationary models that do
not bounce, but inflate from a static beginning, and then reheat
in the usual way.\\

The inflationary universe emerges from a small static state that
has within it the seeds for the development of the microscopic
universe and it is called Emergent Universe scenario. The universe
has a finite initial size, with a finite amount of inflation
occurring over an infinite time in the past and with inflation
then coming to an end via reheating in the standard way. Reference
[8] (see also [11]) summarized the features of emergent universe
as:
\begin{enumerate}
    \item the universe is almost static at the finite past $(t\rightarrow-\infty)$ and isotropic, homogeneous at large
    scales;
    \item it is ever existing and there is no timelike
    singularity;
    \item the universe is always large enough so that the classical
description of space-time is adequate;
    \item the universe may contain exotic matter so that the energy conditions
may be violated;
    \item the universe is accelerating as suggested by recent measurements of distances of
high redshift type Ia supernovae.
\end{enumerate}

Ellis et al [9] provided a realization of a singularity-free
inflationary universe in the form of a simple cosmological model
dominated at early times by a single minimally coupled scalar
field with a physically based potential. Mukherjee et al [8]
presented a general framework for an emergent universe scenario
and showed that emergent universe scenarios are not isolated
solutions and they may occur for different combinations of
radiation and matter. Campo et al [10] studied the emergent
universe model in the context of a self-interacting
Jordan-Brans-Dicke theory and showed that the model presents a
stable past eternal static solution which eventually enters a
phase where the stability of this solution is broken leading to an
inflationary period. Debnath [11] discussed the behaviour of
different stages of the evolution of the emergent universe
considering that the universe is filled with normal matter and a
phantom field. Mukherji and Chakraborty [12] developed
Einstein-Gauss-Bonnet (EGB) theory in the emergent universe
scenario. They have considered the
Friedman-Lemaître-Robertson-Walker cosmological model in Horava
gravity and the emergent scenario for all values of the spatial
curvature. Paul et al [13] predicted the range of the permissible
values for the parameters associated with the constraints on
exotic matter needed for an emergent universe. The emergent
universe in Ho¢rava gravity was studied by Mukherji and
Chakraborty [14].
\\

Many theoretical models that anyhow describing the accelerated
expansion of the universe and which appears to fit all currently
available observations are affected by significant fine-tuning
problems related to the vacuum energy scale and therefore it is
important to investigate alternatives to this description of the
Universe [15]. There exist several other approaches to the
theoretical description of the accelerated expansion of the
universe. One of these is a modified gravity theories. Studies of
the physics of these theories is however hampered by the
complexity of the field equations, making it difficult to obtain
both exact and numerical solutions which can be compared with
observations. These problems can be reduced somewhat by using the
theory of dynamical systems, which provides a relatively simple
method for obtaining exact solutions and a description of the
global dynamics. Modified gravity constitutes an interesting
dynamical alternative to $\Lambda$CDM cosmology in that it is also
able to describe with success the current acceleration in the
expansion of our universe, the present dark energy epoch [16].
Presumably the simplest modified gravity model of dark energy is
so-called $f(R)$ gravity in which $f$ is a function in terms of a
Ricci scalar $R$ ([17], [18], [19], [20], [21], [22]). The $f(R)$
theories do not seem to introduce any new type of matter and can
lead to late time acceleration [23]. Another modified gravity
theory is the Chameleon field theory where the scalar field
interacts with some kind of matter, behaving as perfect fluid. The
Lagrangian of the model contains a term, where the matter
interacts with the effective metric (physical metric, multiplied
by a conformal factor depending on the scalar field)[24]. Brax et
al [23] reviewed both $f(R)$ gravity and Chameleon field theories
and discussed association between them. Another interesting sort
of modified theories is so-called $f(T)$ gravity (where $T$ is the
torsion)([25], [26], [27]). Recently, it is shown that such $f(T)$
gravity theories also admit the accelerated expansion of the
universe without resorting to dark energy ([27], [28]). In the
present paper, we have discussed the emergent universe scenario in
the generalized gravity theories like Chameleon
gravity, $f(R)$ gravity and $f(T)$ gravity.\\

The organization of the paper is as follows: In section II, we
have given the basic field equations in the Einstein's gravity. We
have chosen some particular form of scale factor for emergent
scenario of the universe and made some restrictions upon the
parameters. In section III, we have discussed the Chameleon field
gravity model and shown the behaviour of the potential function
for emergent scenario. In section IV and V, we have studied the
$f(R)$ and $f(T)$ gravity models respectively and shown the nature
of equation of states in emergent scenarios of the universe.
Finally, some conclusions have been drawn in section VI.\\

\section{\bf\large{Basic Equations}}

For a FRW spacetime, the line element is given by

\begin{equation}
ds^{2} = - dt^{2} + a^{2}(t) \left[\frac{ dr^{2}}{1-k r^{2}} +
r^{2}(d\theta^{2}+\sin^{2}\theta d\phi^{2})\right]
\end{equation}

where $a(t)$ is the scale factor and $k ~(= 0, \pm 1)$ is the
curvature scalar. Now consider the Hubble parameter ($H$) and the
deceleration parameter ($q$) in terms of scale factor as

$$
H=\frac{\dot{a}}{a}~~,~~q=-\frac{a\ddot{a}}{\dot{a}^{2}}=-1-\frac{\dot{H}}{H^{2}}
$$

So the Einstein field equations are given by (choosing $c=1$)

\begin{equation}
H^{2}+\frac{k}{a^{2}}=\frac{8\pi G\rho}{3}
\end{equation}
and
\begin{equation}
\dot{H}-\frac{k}{a^{2}}=-4\pi G(\rho+p)
\end{equation}

and the energy conservation equation is

\begin{equation}
\dot{\rho}+3H(\rho+p)=0
\end{equation}

where $\rho$ and $p$ are respectively the density and pressure of
the fluid.\\

For emergent universe, the scale factor is chosen as [8]

\begin{equation}
a=a_{0}\left(\beta+e^{\alpha t}\right)^{n}
\end{equation}

where the constant parameters are restricted as follows [14]:\\
\\
1. $a_{0}>0$ for the scale factor a to be positive, \\
2. $\beta>0$, to avoid any singularity at finite time (big-rip),\\
3. $\alpha>0$ or $n>0$ for expanding model of the universe,\\
4. $\alpha<0$ and $n<0$ implies big bang singularity at
$t=-\infty$.\\

For this choice of the scale factor, the Hubble parameter, its
derivatives are given by

$$
H=\frac{n\alpha e^{\alpha t}}{e^{\alpha
t}+\beta}~,~~~~~~~~~\dot{H}=\frac{n \beta \alpha^{2}e^{\alpha
t}}{(e^{\alpha t}+\beta)^{2}} ~,~~~~~~~~~~~~~~~~\ddot{H}=\frac{n
\beta \alpha^{3}e^{\alpha t}(\beta-e^{\alpha t})}{(e^{\alpha
t}+\beta)^{3}}
$$

Here, $H$, $\dot{H}$, and $\ddot{H}$ tend to $0$ as
$t\rightarrow-\infty$. We have seen that $\dot{H}>0$ for emergent
scenario. From (2) and (3), it can be observed that the emergent
scenario is possible for (i) flat ($k=0$) universe in phantom
stage only, (ii) open ($k=-1$) universe only for phantom stage if
$H>\frac{1}{a}$ and (iii) closed ($k=+1$) universe in quintessence
stage if $\dot{H}<\frac{1}{a^{2}}$ and in phantom stage if
$\dot{H}>\frac{1}{a^{2}}$.\\

Also we calculated the cosmographic parameters [29, 30]

\begin{equation}
\begin{array}{c}
  q=-\frac{1}{a}\frac{d^{2}a}{dt^{2}}H^{-2}\\\\
  j=\frac{1}{a}\frac{d^{3}a}{dt^{3}}H^{-3}\\\\
  s=\frac{1}{a}\frac{d^{4}a}{dt^{4}}H^{-4}\\\\
  l=\frac{1}{a}\frac{d^{5}a}{dt^{5}}H^{-5}\\\\
\end{array}
\end{equation}

which are usually referred to as the deceleration, jerk, snap, and
lerk parameters. After calculating the cosmographic parameters for
the emergent universe we have plotted them against cosmic time $t$
in the figures 1a and 1b. It is observed in figure 1a that the
deceleration parameter is staying at negative level. This
indicates the ever accelerating nature of the emergent universe.
In figure 1b we observe that the snap, jerk and lark parameters
are increasing function of the cosmic time in the emergent
universe.
\\\\

\begin{figure}
\includegraphics[height=1.8in]{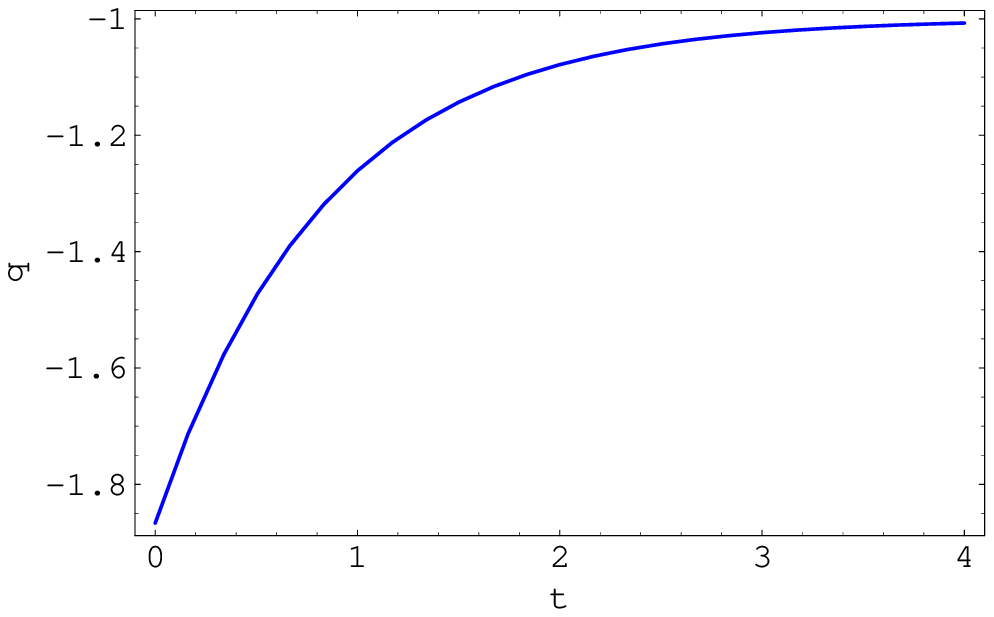}~~~~
\includegraphics[height=1.8in]{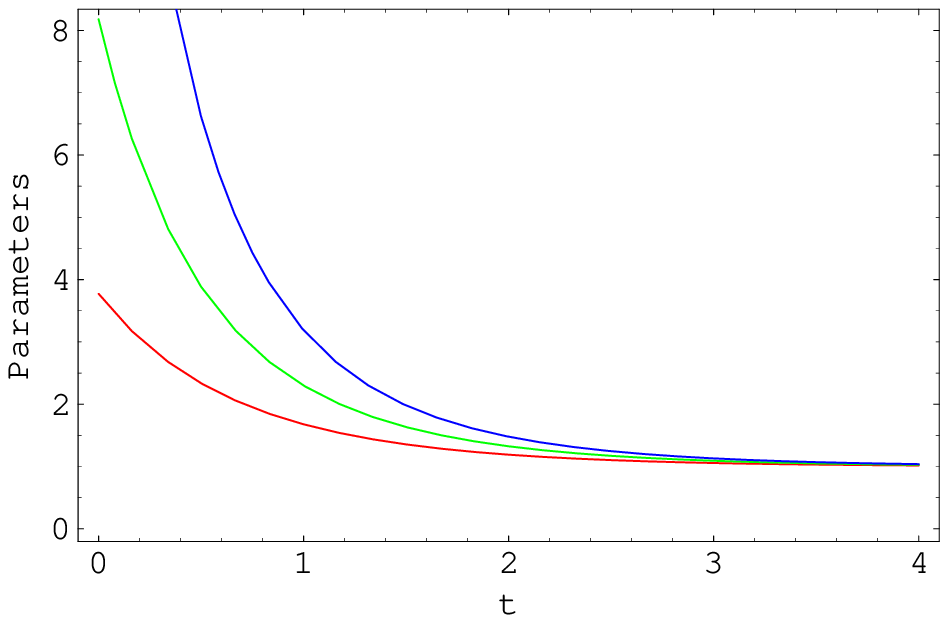}\\
\vspace{1mm}
~~~~~Fig.1a~~~~~~~~~~~~~~~~~~~~~~~~~~~~~~~~~~~~~~~~~~~~~~~~~Fig.1b\\

\vspace{6mm} ~~~~~~~~~~~~Fig.1a  shows the deceleration parameter
$q$ for emergent universe. Fig.1b shows snap parameter $s$ (green line, middle), jerk parameter $j$ (red line, bottom) and lerk parameter $l$ (blue line, top) for emergent universe.\\

\vspace{6mm}

\end{figure}

\section{\bf\large{Emergent scenario in Chameleon gravity}}

In the flat homogeneous universe, we consider that the relevant
action given by [31, 32, 33]

\begin{equation}
S=\int\sqrt{-g}d^{4}x\left[f(\phi)\mathcal{L}+\frac{1}{2}\phi_{,\mu}\phi^{,\mu}+\frac{\mathcal{R}}{16\pi
G}-V(\phi)\right]
\end{equation}

where $\phi$ is the Chameleon scalar field and $V(\phi)$ is the
Chameleon potential. Also, $\mathcal{R}$ is the Ricci scalar and
$G$ is the Newtonian constant of gravity, $f(\phi)\mathcal{L}$ is
the modified matter Lagrangian and $f(\phi)$ is an analytic
function of $\phi$. The variation of action (9) with respect to
the metric tensor components in a FRW cosmology yields the field
equations (choosing $8\pi G=1$),

\begin{equation}
H^{2}=\frac{1}{3}\left[\rho
f+\frac{1}{2}\dot{\phi}^{2}+V(\phi)\right]-\frac{k}{a^{2}}
\end{equation}
and
\begin{equation}
\dot{H}=\frac{1}{2}\left[-pf-\frac{1}{2}\dot{\phi}^{2}+V(\phi)\right]+\frac{k}{a^{2}}
\end{equation}

The conservation equation and the wave equation in presence of Chameleon field are given by [31]

\begin{equation}
\frac{\partial}{\partial t}(\rho
f)+3H(p+\rho)f=(p+\rho)\dot{f}
\end{equation}
and
\begin{equation}
3H\dot{\phi}^2+\dot{\phi}\ddot{\phi}+\dot{V}+(p+\rho)\dot{f}=0
\end{equation}

\begin{figure}
\includegraphics[height=2.0in]{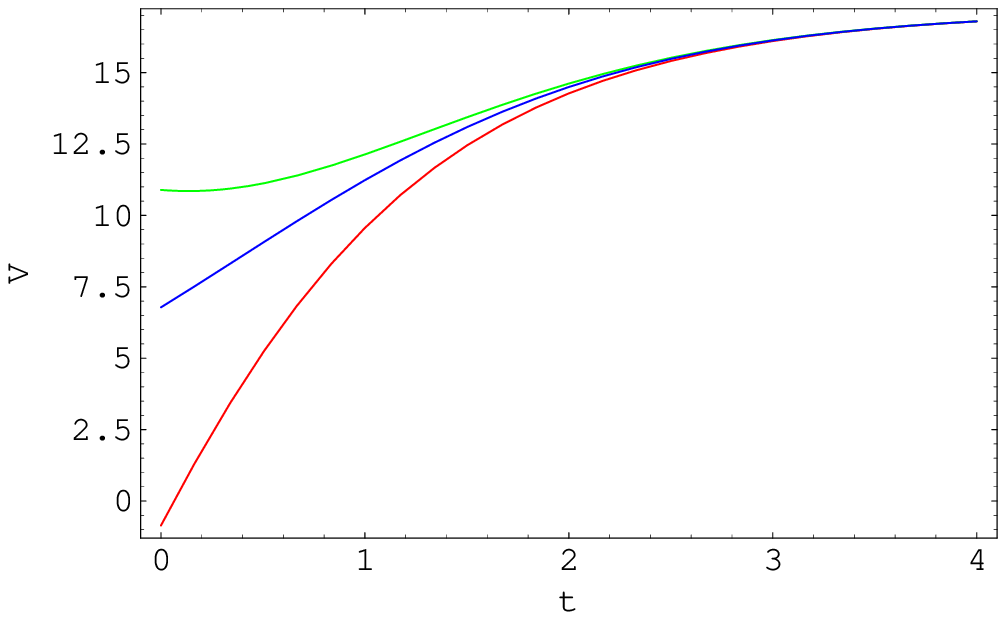}~\\
\vspace{1mm} ~~~~~~~~~~~~Fig.2a. This figure shows the variation
of potential $V$ of the Chameleon field with cosmic time $t$ in
the emergent universe scenario. The cases of flat $(k=0)$ (the
blue line, middle), open $(k=-1)$ (the red line, bottom) and
closed $(k=1)$ (the green line, top) universe are considered.

\vspace{6mm}

\end{figure}

\begin{figure}
\includegraphics[height=1.8in]{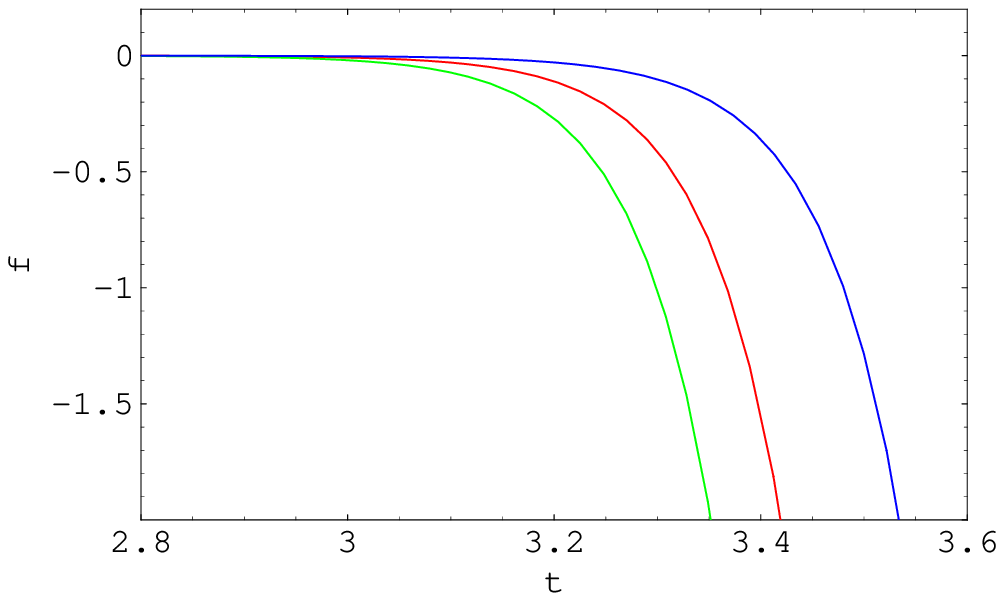}~~~~
\includegraphics[height=1.8in]{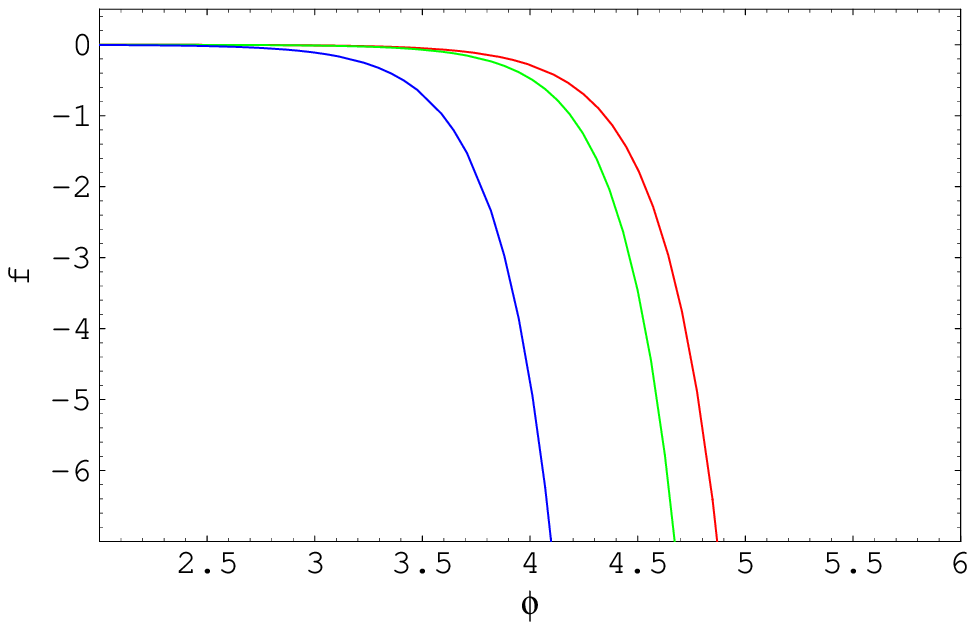}\\
\vspace{1mm}
~~~~~Fig.2b~~~~~~~~~~~~~~~~~~~~~~~~~~~~~~~~~~~~~~~~~~~~~~~~~Fig.2c\\

\vspace{6mm} ~~~~~~~~~~~~Fig.2b  shows the variation of the
analytic function $f$ of the Chameleon field with cosmic time $t$
in the emergent universe scenario. In this figure, the cases of
flat $(k=0)$ (the blue line, top), open $(k=-1)$ (the red line,
middle) and closed $(k=1)$ (the green line, bottom) universe are
considered. Fig.2c shows the variation of the analytic function
$f$ of the Chameleon field with $\phi$ in the emergent universe
scenario. In this figure, the cases of flat $(k=0)$ (the blue
line, bottom), open $(k=-1)$ (the red line, top)
and closed $(k=1)$ (the green line, middle) universes are considered.\\

\vspace{6mm}

\end{figure}

\begin{figure}
\includegraphics[height=1.8in]{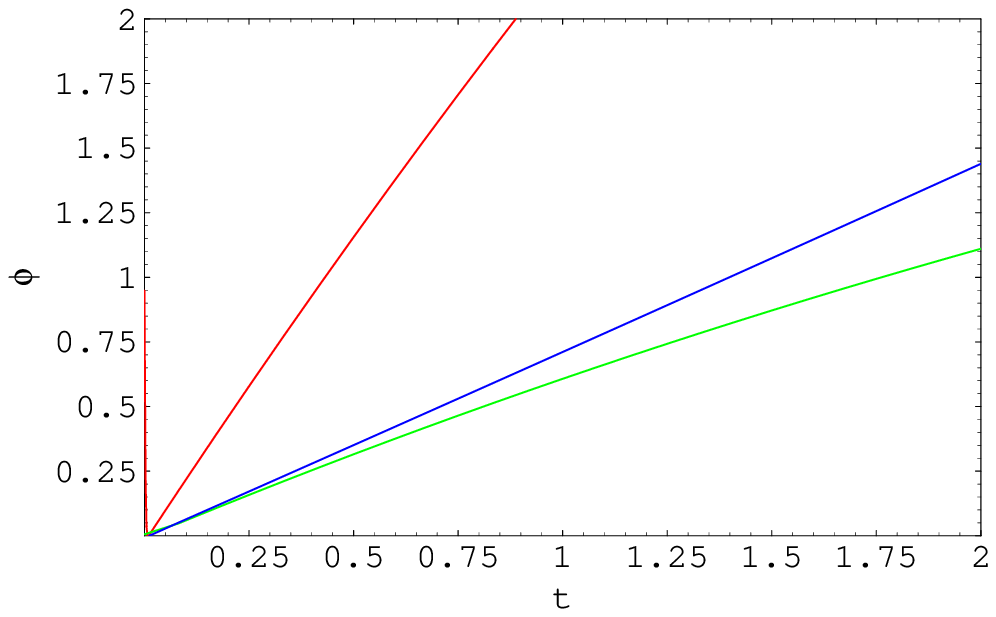}\\
\vspace{1mm}~~~~~~~~~~~~~~~Fig.3\\

\vspace{6mm}

Fig. 3 represents the evolution of the scalar field $\phi$ of the
Chameleon field with time $t$ in the emergent universe scenario
for open $(k=-1)$ (the red line, top), closed $(k=1)$ (the green
line, bottom) and flat $(k=0)$ (the blue line, middle), universe
respectively.

 \vspace{7mm}

\end{figure}

\begin{figure}
\includegraphics[height=1.8in]{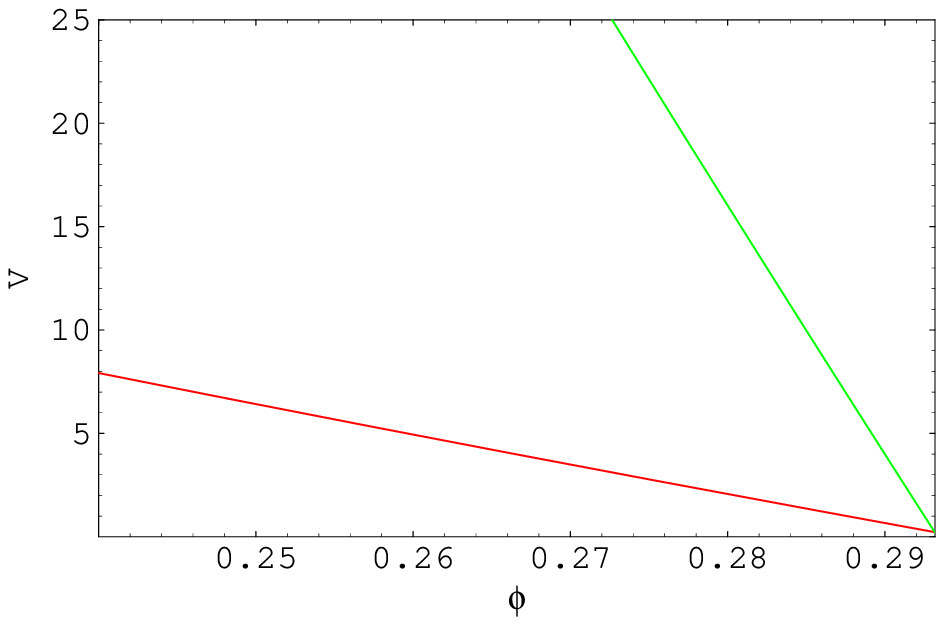}~~~~
\includegraphics[height=1.8in]{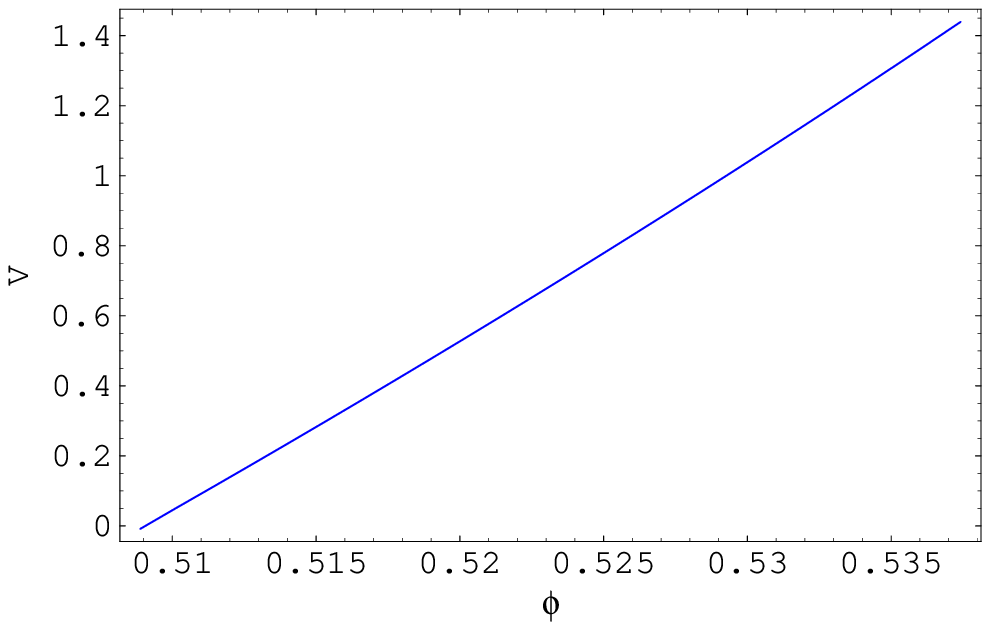}\\
\vspace{1mm}
~~~~~Fig.4a~~~~~~~~~~~~~~~~~~~~~~~~~~~~~~~~~~~~~~~~~~~~~~~~~Fig.4b\\

\vspace{6mm}

Fig. 4a shows the potential $V$ against the scalar field $\phi$ in
the cases of $(k=-1)$ (the red line, bottom) and closed $(k=1)$
(the green line, top). Fig. 4b shows the potential $V$ against the
scalar
field $\phi$ in the case of flat $k=0$ universe.\\

\vspace{7mm}

\end{figure}

The early universe contains a standard scalar field $\phi$ with
energy density $\rho_{\phi}=\frac{1}{2}\dot{\phi}^{2}+V(\phi)$ and
pressure $p_{\phi}=\frac{1}{2}\dot{\phi}^{2}-V(\phi)$ and possibly
also ordinary matter with energy density $\rho$ and pressure
$p=w\rho$, where $-\frac{1}{3}\leq w \leq 1$. The cosmological
constant is absorbed into the potential $V$. There are no
interactions between matter and the scalar field. Following
references [31,32] we make the simplifying assumption

\begin{equation}
V=V_{0}\dot{\phi}^{2}
\end{equation}

where, $V_{0}$ is a positive constant. Using this simplifying
assumption in the Chameleon field equation along with the
specified form of scale factor (5) and equations (8) and (9), we get (after simplification) the form of $\phi$ in
the emergent scenario as

\begin{equation}
\phi=\int\sqrt{\frac{2(k(1+3kw)(e^{\alpha
t}+\beta)^{-2n}+e^{\alpha t}n\alpha^{2}(e^{\alpha
t}+\beta)^{-2}(3e^{\alpha
t}n(1+w)+2\beta)a_{0}^{2})}{a_{0}^{2}(-1+w+2(1+w)V_{0})}}dt
\end{equation}

Using (5), (8), (12) and (13) and the $\rho=\rho_{0}a^{-3(1+w)}f^{w}$ (from (10)) for
Chameleon field from reference [31], we express $f(\phi)$ as a
function of $t$ as follows

\begin{equation}
f=\left[\frac{((e^{t\alpha}+\beta)^{n}a_{0})^{1+3w}(k(e^{t\alpha}+\beta)^{2}(-4+3(1-k)w+(4+6(1-k)w)V_{0})-2e^{t\alpha}n\alpha^{2}(e^{t\alpha}+\beta)^{2n}a_{0}^{2}(3e^{t\alpha}n+\beta(1+2V_{0})))}{((e^{t\alpha}+\beta)^{2}(-1+w+2(1+w)V_{0})\rho_{0})}\right]^{\frac{1}{1+w}}
\end{equation}

We see that $(e^{\alpha t}+\beta)^{-2}(3e^{\alpha
t}n(1+w)+2\beta)a_{0}^{2})$ is always positive. Thus from (13) we
see that emergent scenario is possible for flat, open and closed
universe under the following conditions:

\begin{itemize}
    \item In the case of $k=0$, $\dot{\phi}^{2}>0$ for $V_{0}>\frac{1-w}{2(1+w)}$
    \item In the case of $k=-1$, $\dot{\phi}^{2}>0$ for
    $w>\frac{1-e^{\alpha t}n\alpha^{2}(e^{\alpha t}+\beta)^{2(-1+n)}(3ne^{\alpha t}
    +2\beta)a_{0}^{2}}{3(1+e^{2\alpha t}n^{2}\alpha^{2}a_{0}^{2}(e^{\alpha t}+\beta)^{2(-1+n)})}$
    and $V_{0}>\frac{1-w}{2(1+w)}$
    \item In the case of $k=1$, $\dot{\phi}^{2}>0$ for
    $w>-\frac{1+e^{\alpha t}n\alpha^{2}(e^{\alpha t}+\beta)^{2(-1+n)}(3ne^{\alpha t}
    +2\beta)a_{0}^{2}}{3(1+e^{2\alpha t}n^{2}\alpha^{2}a_{0}^{2}(e^{\alpha t}+\beta)^{2(-1+n)})}$ and
    $V_{0}>\frac{1-w}{2(1+w)}$.
\end{itemize}

Thus, under the following conditions we consider the cases of flat
$(k=0)$, open $(k=-1)$ and closed $(k=+1)$ universe. In figure 2a,
we plot the potential against cosmic time. We find that the
potential is increasing with time in all of the three cases. While
plotting, the constant parameters are chosen according to the
above four constraints. Evolution of $f(\phi)$ is presented in
figures 2b and 2c. In all of the three situations $f(\phi)$ is
staying at negative level and is showing a falling behaviour with
cosmic time as well as the scalar field $\phi$. For open $(k=-1)$,
closed $(k=+1)$, and flat $(k=0)$ universe, the evolution of the
scalar field $\phi$ exhibits an increasing pattern with evolution
of the universe. These are presented in figure 3 where the scalar
field $\phi$ is plotted against cosmic time $t$. Figure 4a shows
that potential $V$ is decreasing with the scalar field $\phi$ for
$k=-1$ and $k=+1$. However, figure 4b shows that for $k=0$ i.e.
for open universe, $V$ is increasing with $\phi$.\\\\

\section{\bf\large{Emergent scenario in} $f(R)$ {gravity}}

Recently, motivated by astrophysical data which indicate that the
expansion of the universe is accelerating, the modified theory of
gravity (or $f(R)$ gravity) which can explain the present
acceleration without introducing dark energy, has received intense
attention [34-40]. In a recent work, Nojiri and Odinstov [38]
suggested two realistic $f(R)$ and one $F(G)$ modified gravities
characterized by the presence of the effective cosmological
constant epochs in such a way that early-time inflation and
late-time cosmic acceleration are naturally unified within single
model. In another work, Nojiri and Odintsov [39] proposed another
class of modified $f(R)$ gravity which unifies $R^{m}$ inflation
with $\Lambda$CDM era. Cognola et al [40] demonstrated a class of
viable modified $f(R)$ gravities describing inflation and the
onset of accelerated expansion. In $f(R)$ gravity action is
described by an arbitrary function of the scalar curvature $R$.
Extensive review of $f(R)$ gravity is available in Nojiri and
Odintsov [21, 22]. The action of $f(R)$ gravity is given by
[44-47]

\begin{equation}
S=\int
d^{4}x\sqrt{-g}\left[\frac{f(R)}{2\kappa^{2}}+\mathcal{L}_{matter}\right]
\end{equation}
where $g$ is the determinant of the metric tensor $g_{\mu\nu}$,
$\mathcal{L}_{matter}$ is the matter Lagrangian and
$\kappa^{2}=8\pi G$. The$f(R)$ is a non-linear function of the
Ricci curvature $R$ that incorporates corrections to the
Einstein-Hilbert action which is instead described by a linear
function $f(R)$. The gravitational field equations in this theory
are

\begin{equation}
H^{2}=\frac{\kappa^{2}}{3f'(R)}(\rho+\rho_{c})
\end{equation}

\begin{equation}
\dot{H}=-\frac{\kappa^{2}}{2f'(R)}(\rho+p+\rho_{c}+p_{c})
\end{equation}

where $\rho_{c}$ and $p_{c}$ can be regarded as the energy density
and pressure generated due to the difference of $f(R)$ gravity
from general relativity given by [45]

\begin{equation}
\rho_{c}=\frac{1}{\kappa^{2}}\left[\frac{1}{2}(-f(R)+Rf'(R))-3H\dot{R}f''(R)\right]
\end{equation}

\begin{equation}
p_{c}=\frac{1}{\kappa^{2}}\left[\frac{1}{2}(f(R)-Rf'(R))+(2H\dot{R}+\ddot{R})f''(R)+\dot{R}^{2}f'''(R)\right]
\end{equation}

where, the scalar tensor $R=6(\dot{H}+2H^{2})$ [39,45]. Here,
$H=\frac{\dot{a}}{a}$ is the Hubble parameter. Here, $\rho$ and
$p$ are the matter–energy density and pressure respectively. In
the spatially flat universe, i.e. $k=0$, with the matter as dust,
namely $p=0$, following Feng [45],
$p=3H_{0}^{2}\Omega_{m0}e^{-3x}$ with $x=ln
a,~~\Omega_{m0}=\rho_{m0}/3H_{0}^{2}$ and $H_{0}$ is the present
Hubble parameter. Following references [36] we choose $f(R)$ as

\begin{equation}
f(R)=R+\xi R^{\mu}+\zeta R^{-\nu}
\end{equation}

where, constants $\mu>0;~~\nu>0$. Subsequently, we get from
equation (17)

\begin{equation}
\begin{array}{c}
\dot{H}=(a^{3}HR(\zeta\nu(1+\nu)+R^{\mu+\nu}(-1+\mu)\mu\xi)\dot{R}+a^{3}(\zeta\nu(1+\nu)(2+\nu)-R^{\mu+\nu}(-2+\mu)(-1+\mu)\mu\xi)\dot{R}^{2}+\\
 a^{3}R(-\zeta\nu(1+\nu)-R^{\mu+\nu}(-1+\mu)\mu\xi)\ddot{R}-3R^{3+\nu}\kappa^{2}H_{0}^{2}\Omega_{m0})\times\\
 (2a^{3}R^{2}(-\zeta\nu+R^{\nu}(R+R^{\mu}\mu\xi)))^{-1} \\
\end{array}
\end{equation}

As both $R$ and $H$ are functions of the scale factor $a$ and its
derivatives, it seems that the equation (21) is set for yielding
the solution for the scale factor. However, it involves
fourth-order derivatives of $a$ ($R$ already contains $\ddot{a}$)
and is highly nonlinear. This makes it difficult to obtain a
completely analytic solution for $a$. Mukherjee et al [8] obtained
the general solution of the scale factor for emergent universe in the form presented in equation (6) without
referring to the actual source of the energy density. Se we use
equation (6) as the choice of scale factor for emergent nature of the universe and reconstruct
$\rho_{c}$ and $p_{c}$ based on $R$ in the emergent universe and
are expressed as follows

\begin{equation}
\begin{array}{c}
  \rho_{c}=\frac{1}{(2ne^{t\alpha}+\beta)^{2}\kappa^{2}}(2^{-1-\nu}3^{-\nu}\varrho^{-\nu}((1+\nu)(-4e^{2t\alpha}n^{2}-e^{t\alpha}(-\nu+4n(1+\nu))\beta-(1+\nu)\beta^{2})\zeta\\
  +\varrho^{\mu+\nu})(2^{2+\mu+\nu}3^{\mu+\nu}e^{2t\alpha}n^{2}(-1+\mu)-6^{\mu+\nu}e^{t\alpha} \\
  \times(4n(-1+\mu)-\mu)(-1+\mu)\beta+(2^{1+\mu+\nu}3^{\mu+\nu}\mu-6^{\mu+\nu}(1+\mu^{2}))\beta^{2})\xi)) \\
\end{array}
\end{equation}

\begin{equation}
\begin{array}{c}
 p_{c}=\frac{1}{6\kappa^{2}}(\varrho^{-\nu}(2^{-\nu}3^{1-\nu}((1+\nu)\zeta-6^{\mu+\nu}(-1+\mu)\xi\varrho^{\mu+\nu}))+\\
  \frac{1}{n\alpha(2ne^{t\alpha}+\beta)^{2}}(6^{-\nu}e^{-t\alpha}\beta(e^{t\alpha}(-1+4n)+\beta)(e^{t\alpha}(1+2n\alpha)+\beta)\times\\
  +(\nu(1+\nu)\zeta+6^{\mu+\nu}(-1+\mu)\mu\xi\varrho^{\mu+\nu}))+\frac{1}{n(2ne^{t\alpha}+\beta)^{3}}(6^{-\nu}e^{-t\alpha}\beta^{2}(e^{t\alpha}(-1+4n)+\beta)^{2}(-\nu(1+\nu)(2+\nu)\zeta+ \\
  +6^{\mu+\nu}(-2+\mu)(-1+\mu)\mu\xi\zeta^{\mu+\nu}\\
\end{array}
\end{equation}

where

\begin{equation}
\varrho=\frac{n\alpha^{2}e^{t\alpha}(2ne^{t\alpha}+\beta)}{(e^{t\alpha}+\beta)^{2}}
\end{equation}

The $\rho_{c}$ and $p_{c}$ are plotted in figures 5a and 5b
respectively. The figures show that although $\rho_{c}$ is
increasing with cosmic time $t$, the negative $p_{c}$ is falling
with cosmic time. While summarizing the attractiveness of the
modified gravity approach, Nojiri and Odintsov [35] stated that
"modified gravity may naturally describe the transition from
non-phantom phase to phantom one without necessity to introduce
the exotic matter". Abdalla et al [49] had shown in an earlier
study that the modified gravity minimally coupled with the usual
(or quintessence) matter may reproduce the quintessence (or
phantom) evolution phase for the dark energy universe in an easier
way than without such coupling. In a more recent study, Bamba et
al [50] studied a viable model of modified gravity in which the
transition from the de Sitter universe to the phantom phase can
occur. Motivated by these earlier studies we examine the bahaviour
of the equation of state parameter in $f(R)$ gravity. The equation
of state parameter plotted in figure 6 is found to stay above
$-1$. This indicates quintessence behaviour. This means that
evolution from quintessence from phantom is not available in the
case of $f(R)$ gravity in emergent universe. From figure 7 we find
the increasing behavior of $f(R)$ with the evolution of the
universe.
\\\\

\begin{figure}
\includegraphics[height=1.8in]{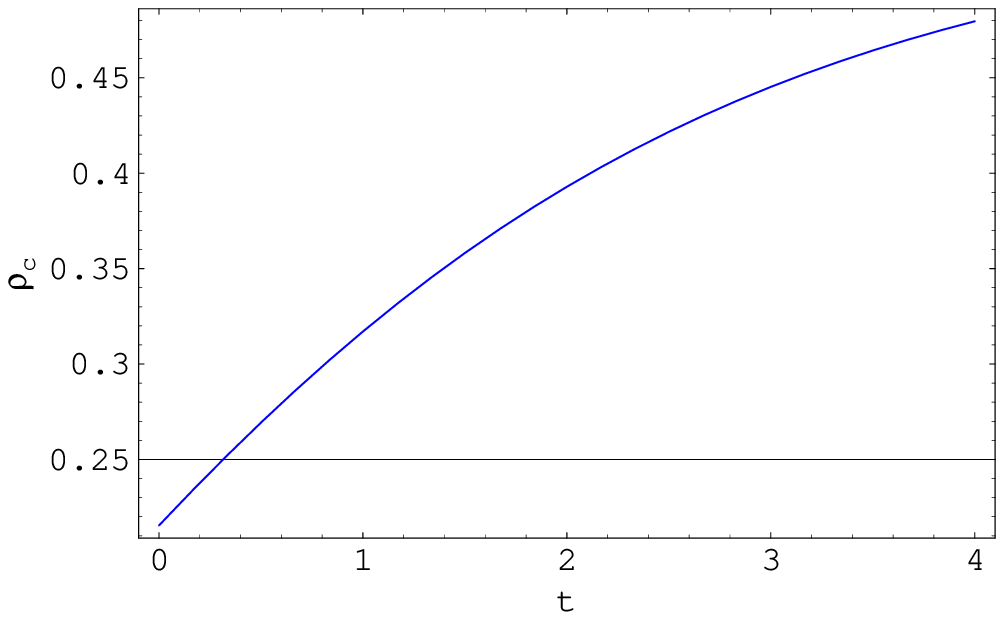}~~~~
\includegraphics[height=1.8in]{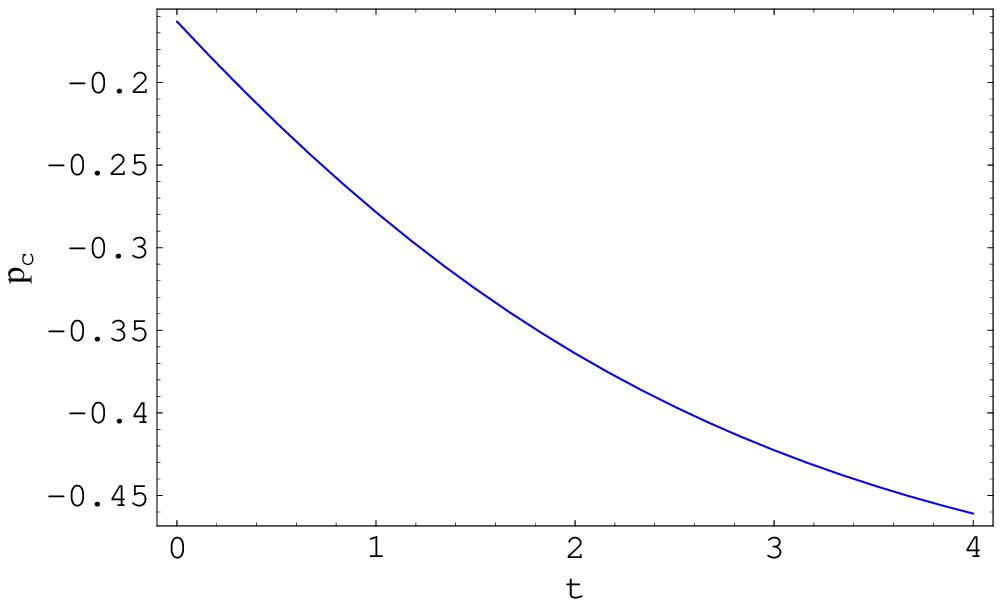}\\
\vspace{1mm}
~~~~~Fig.5a~~~~~~~~~~~~~~~~~~~~~~~~~~~~~~~~~~~~~~~~~~~~~~~~~~~~~~~~~~~~Fig.5b\\

\vspace{6mm}

Figs. 5a and 5b represent the evolution of the energy density
$\rho_{c}$ and pressure $p_{c}$ generated due to the difference of
$f(R)$ gravity from general relativity.

\vspace{7mm}

\end{figure}

\begin{figure}
\includegraphics[height=1.8in]{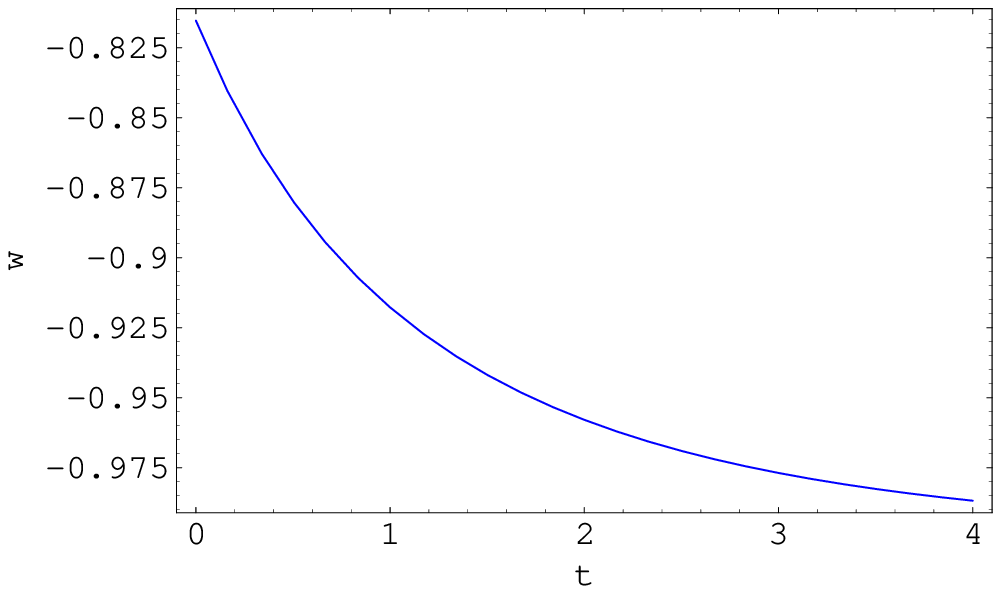}~~~~
\includegraphics[height=1.8in]{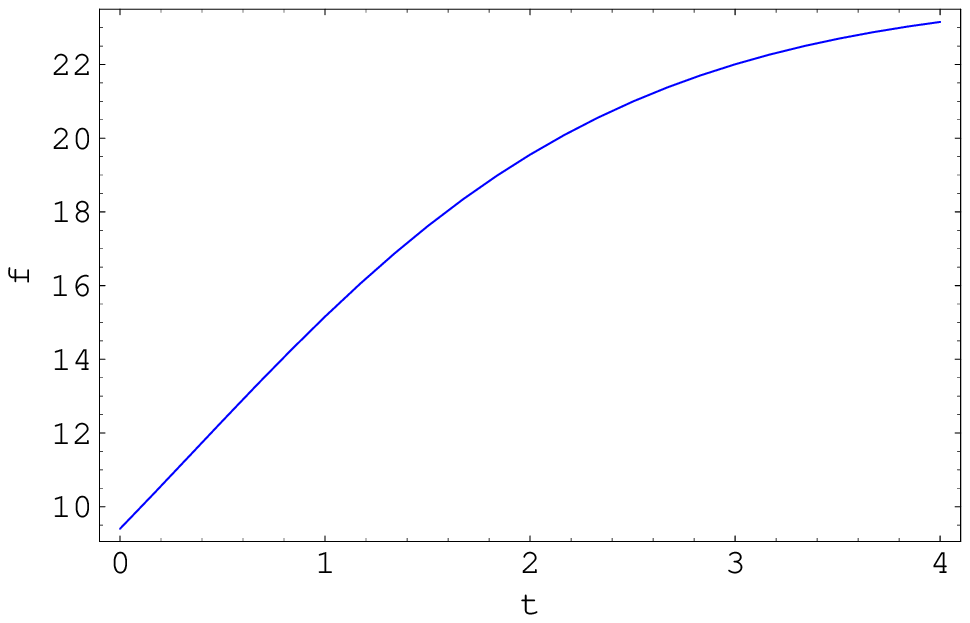}~\\
\vspace{1mm}
~~~~~Fig.6~~~~~~~~~~~~~~~~~~~~~~~~~~~~~~~~~~~~~~~~~~~~~~~~~~~~~~~~~~~~~~Fig.7

\vspace{6mm} Fig.6 shows the evolution of the equation of state
$w_{c}=\frac{p_{c}}{\rho_{c}}$ with cosmic time $t$.\\ Fig.7
represents the evolution of $f(R)$ with cosmic time $t$.

\vspace{6mm}

\end{figure}

\section{\bf\large{Emergent scenario in} $f(T)$ {gravity}}

Recently, models based on modified teleparallel gravity were
presented as an alternative to inflationary models. The theory so
obtained is called as the $f(T)$ theory [51-57]. In this theory
instead the curvature defined via the Levi-Civita connection, the
so-called Weitzenbock connection is used. But in this case the
theory has no curvature but instead torsion. Similar to general
relativity where the action is the curvature scalar, the action of
teleparallel gravity is a torsion scalar $T$ ([51], [52], [58] ).
The action of $f(T )$ theory is obtained by replacing $T$ in the
action of teleparallel gravity by $T+f(T)$ [53]. We start with the
following action for the $f(T)$ gravity [52]

\begin{equation}
S=\frac{1}{2\kappa^{2}}\int
d^{4}x\sqrt{-g}[T+f(T)+\mathcal{L}_{m}]
\end{equation}

where $T$ is the torsion scalar, $f(T)$ is general differentiable
function of the torsion and $\mathcal{L}_{m}$ corresponds to the
matter Lagrangian and $\kappa^{2}=8\pi G$. The torsion scalar $T$
is defined as

\begin{equation}
T=S_{\rho}^{\mu\nu}T^{\rho}_{\mu\nu}
\end{equation}
with
\begin{equation}
S_{\rho}^{\mu\nu}=\frac{1}{2}(K^{\mu\nu}_{\rho}+\delta_{\rho}^{\mu}T^{\theta\mu}_{\theta}-\delta_{\rho}^{\mu}T^{\theta\mu}_{\theta})
\end{equation}

where, $K^{\mu\nu}_{\rho}$ is the contorsion tensor, which equals
the difference between Weitzenbock and Levi-Civita connections. So
$f(T)$ gravity uses the curvatureless Weitzenbock connection.\\

\begin{figure}
\includegraphics[height=1.8in]{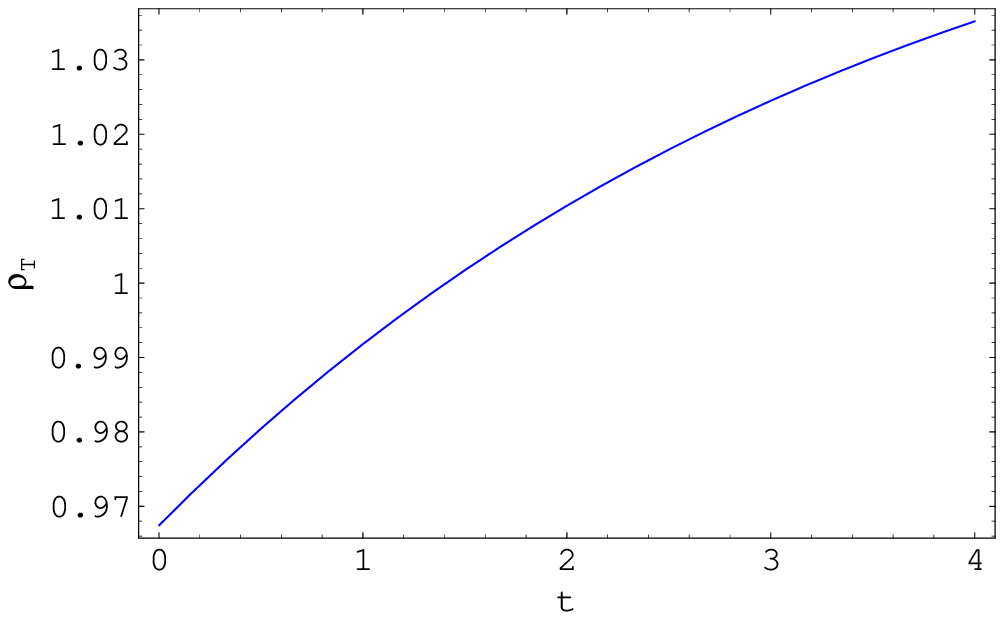}~~~~
\includegraphics[height=1.8in]{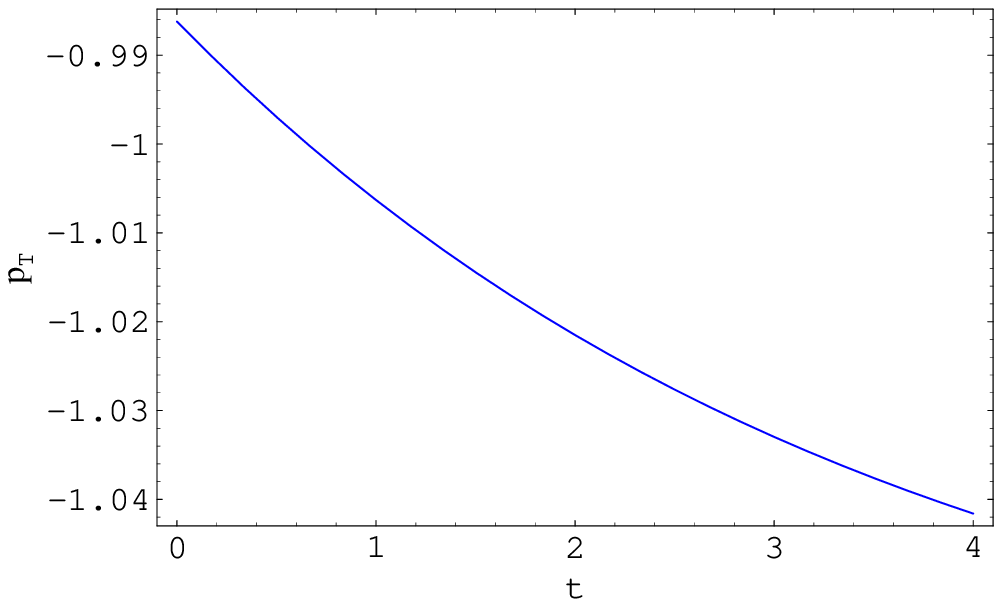}\\
\vspace{1mm}
~~~~~Fig.8a~~~~~~~~~~~~~~~~~~~~~~~~~~~~~~~~~~~~~~~~~~~~~~~~~~~~~~~~~~~Fig.8b\\

\vspace{6mm}

Figs. 8a and 8b represent the evolution of the energy density
$\rho_{T}$ and pressure $p_{T}$ respectively against cosmic time
$t$.

\vspace{7mm}

\end{figure}

\begin{figure}
\includegraphics[height=1.8in]{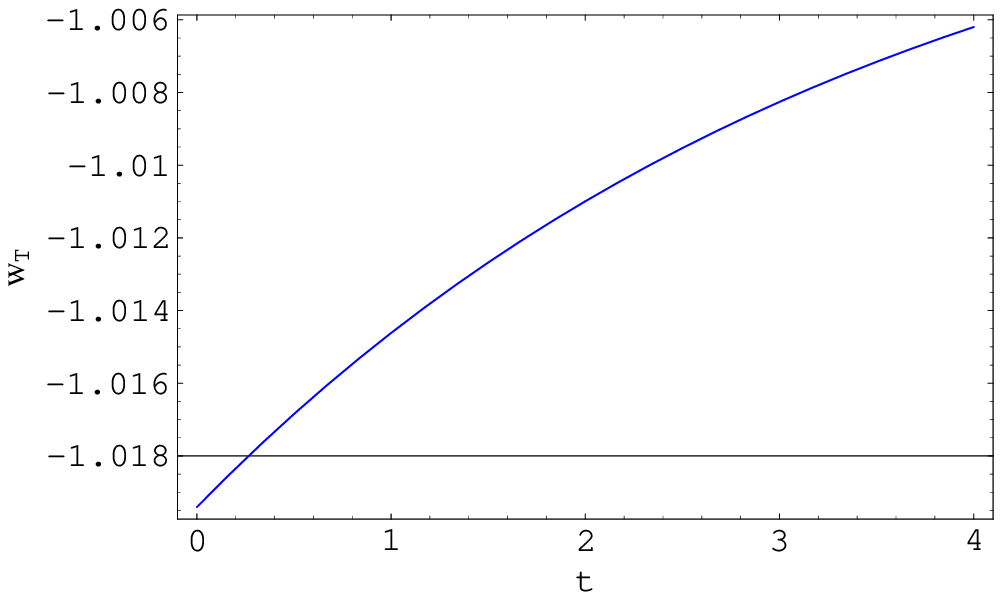}~~~~
\includegraphics[height=1.8in]{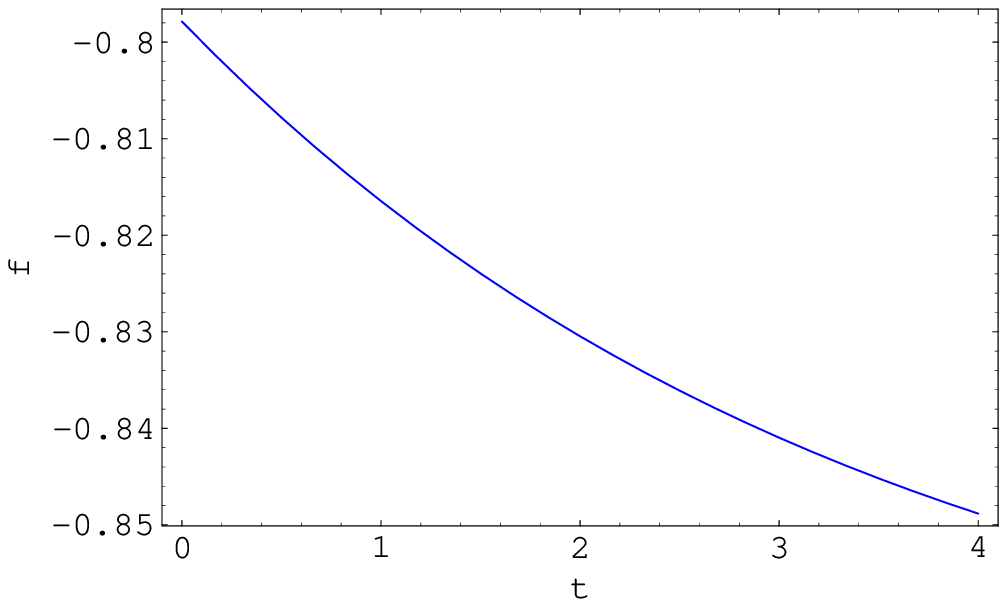}\\
\vspace{1mm}
~~~~~Fig.9~~~~~~~~~~~~~~~~~~~~~~~~~~~~~~~~~~~~~~~~~~~~~~~~~~~~~~~~~~~~~Fig.10

\vspace{6mm}

Fig. 9 shows the evolution of the equation of state
$w_{T}=\frac{p_{T}}{\rho_{T}}$ with cosmic time $t$.\\ Figs. 10
represents the evolution of $f(T)$ with cosmic time $t$.

\vspace{6mm}

\end{figure}

The modified Friedmann equations of motion are ([27], [56])

\begin{equation}
H^{2}=\frac{8\pi G}{3}\rho_{m}-\frac{f(T)}{6}-2f'(T) H^{2}
\end{equation}

\begin{equation}
\dot{H}=-\frac{4\pi G(\rho_{m}+p_{m})}{1+f'(T)-12H^{2}f''(T)}
\end{equation}

where, $\rho_{m}$ and $p_{m}$ respectively for the energy density
and pressure of the matter content of the universe, with
equation-of-state parameter $w_{m}=p_{m}/\rho_{m}$. Since we are
interested in the late time universe, we consider only
non-relativistic matter (cold dark matter and baryon), we have
$p_{m}=0$. By comparing the above modified Friedmann equations
 with the ordinary ones in general relativity one
can obtain the energy densities and pressure for $f(T)$ gravity as
[26]

\begin{equation}
\rho_{T}=\frac{1}{2\kappa^{2}}(2Tf'(T)-f(T)+6H^{2})
\end{equation}
and
\begin{equation}
p_{T}=-\frac{1}{2\kappa^{2}}\left[-8\dot{H}Tf''(T)+(2T-4\dot{H})f'(T)-f(T)+4\dot{H}+6H^{2}\right]
\end{equation}

and the equation of state

\begin{equation}
w_{T}=-1-\frac{8\dot{H}Tf''(T)+4\dot{H}f'(T)}{f(T)-2Tf'(T)}
\end{equation}

As we are considering the emergent universe scenario, we use the
scale factor $a$ as in equation (6). Subsequently, $H$ gets the
form in (7). Following reference [49] we take $f(T)$ as

\begin{equation}
f(T)=\gamma T+\lambda T^{m}
\end{equation}

where, $m>0$ and $T=-6H^{2}$. Consequently,

\begin{equation}
\dot{H}=-\frac{36
H^{2}\kappa^{2}H_{0}^{2}\Omega_{m0}}{a^{3}(6H^{2}(1+\gamma)-(-H^{2})^{m}(6^{m}+2^{1+m}3^{m}(-1+m))m\lambda)}
\end{equation}

The above equation is non-linear in $H$ and hence it is difficult
to get an analytical solution of $a$. Thus, like the $f(R)$
gravity we take the scale factor in the form (6) to consider the
emergent universe scenario. Using the above forms of $a$ and
$f(T)$ we reconstruct $\rho_{T}$, $p_{T}$ and the equation of
state parameter $w_{T}$. Forms of $\rho_{T}$ and $p_{T}$ are
obtained as follows

\begin{equation}
\begin{array}{c}
 \rho_{T}=\frac{\kappa^{2}}{2}[-6^{m}\eta^{m}\lambda+\frac{1}{(e^{t\alpha}+\beta)^{5}}(6e^{2t\alpha}n^{2}\alpha^{2}(e^{3t\alpha}+3e^{2t\alpha}\beta+3e^{t\alpha}\beta^{2}+\beta^{2}\\
  +e^{3t\alpha}\gamma+3e^{2t\alpha}\beta\gamma+24e^{2t\alpha} n^{2}\alpha^{3}\beta\gamma)\\
  +(3e^{t\alpha}+\beta)\beta^{2}\gamma-2^{2+m}3^{m}m\alpha\beta(e^{t\alpha}+\beta)^{2}\lambda\eta)]\\
\end{array}
\end{equation}

\begin{equation}
\begin{array}{c}
  p_{T}=-\frac{\kappa^{2}}{2}[\frac{2e^{t\alpha}n\alpha^{2}(ne^{t\alpha}(3+2\gamma)+2\beta)}{(e^{t\alpha}+\beta)^{2}}-6\gamma\eta+\frac{1}{(e^{t\alpha}+\beta)^{5}}\\
  \times(8e^{t\alpha}n\alpha^{3}\beta(3ne^{t\alpha}+\beta)(6e^{2t\alpha}n^{2}\alpha^{2}\gamma-m\eta(6^{m}e^{2t\alpha}+2^{1+m}3^{m}e^{t\alpha}\beta+6^{m}\beta^{2})\gamma))\\
  +\frac{1}{(e^{t\alpha}+\beta)^{8}}(96e^{3t\alpha}n^{3}\alpha^{6}\beta^{2}(6e^{2t\alpha}n^{2}\alpha^{2}(e^{t\alpha}-2\beta)\gamma+m\eta(-6^{m}e^{3t\alpha}\\
  +2^{1+m}3^{m}e^{2t\alpha}(-1+m)\beta+6^{m}e^{t\alpha}(-1+4m)\beta^{2}+2^{1+m}3^{m}m\beta^{3})\lambda))]\\
\end{array}
\end{equation}

where,

\begin{equation}
\eta=-\left[\frac{n^{2}\alpha^{2}e^{2t\alpha}}{(e^{t\alpha}+\beta)^{2}}\right]^{m}
\end{equation}

Figures 8a and 8b show the variation of $\rho_{T}$ and $p_{T}$
with cosmic time $t$. Which show that in the emergent scenario
under $f(T)$ gravity the energy density increases with time, while
the negative pressure falls with increase in the cosmic time. The
equation of state parameter $w_{T}<-1$ as plotted against cosmic
time in figure 9. This indicates the phantom scenario. In figure
10 we find the evolution
of $f(T)$ with cosmic time.\\

\section{\bf\large{Concluding remarks}}

In this work, we have considered emergent universe in generalized
gravity theories like Chameleon, $f(R)$ and $f(T)$ gravities.
While considering the Chameleon field, we have considered flat,
open and closed universes. We have reconstructed the potential of
Chameleon field under emergent universe scenario and observed its
increasing nature with evolution of the universe. Also, we
observed the increasing behavior of the associated scalar field in
all of the flat, open and closed universes. We have derived the
conditions for emergent scenario in the situations of flat, open
and closed universes. Next, we have considered the emergent
universe scenario under $f(R)$ gravity. Taking a particular form
of $f(R)$ we have considered the energy density, pressure and
subsequently the equation of state parameter. It has been revealed
that the equation of state parameter exhibits quintessence like
behaviour (above $-1$). However, considering the behaviour of the
equation of state parameter under $f(T)$ gravity in the emergent
universe scenario, we have seen it to lie below $-1$.
This indicates phantom like behaviour.\\\\

\section{\bf\large{Acknowledgement}}
The authors sincerely acknowledge the warm hospitality provided by
IUCAA, Pune, India, where the work was carried out during a
scientific visit. Thanks are also due to the anonymous reviewer
for giving constructive comments to revise the manuscript.
\\\\

{\bf References:}\\
\\
1.  D. J. Mulryne, R. Tavakol, J. E. Lidsey and G. F. R. Ellis,
{\it Phys. Rev. D} {\bf 71} 123512 (2005).\\\\
2.  M. Gasperini and G. Veneziano (2003), {\it Phys. Rep.} {\bf
373},1.\\\\
3.  P. J. Steinhardt and N. Turok (2002), {\it Phys. Rev. D} {\bf
65},
126003.\\\\
4.  G. F. R. Ellis and R. Maartens, {\it Class. Quantum Grav.}
{\bf 21} 223 (2004).\\\\
5.  G. F. R. Ellis, J. Murugan and C. G. Tsagas, {\it Class.
Quantum Grav.} {\bf 21} 233 (2004).\\\\
6.  A. Borde and A. Vilenkin (1994), {\it Phys. Rev. Lett.} {\bf
72},3305.\\\\
7.  C. Molina-Paris and M. Visser (1999), {\it Phys. Lett. B} {\bf
455},
90.\\\\
8.  S. Mukherjee, B. C. Paul, N. K. Dadhich, S. D. Maharaj and A.
Beesham (2006) {\it Class. Quantum Grav.} {\bf 23} 6927.\\\\
9.  G. F. R. Ellis, J. Murugan and C. G Tsagas (2004) {\it Class.
Quantum Grav.} {\bf 21} 233.\\\\
10. S. del Campo et al (2007) {\it JCAP} {\bf 11} 030.\\\\
11. U. Debnath (2008), {\it Class. Quantum Grav.} {\bf 25} 205019.\\\\
12. S. Mukerji and S. Chakraborty (2010), {\it Int. J. Theor.
Phys.} {\bf 49}
2446.\\\\
13. B. C. Paul et al(2010), {\it Mon. Not. R. Astron. Soc.} {\bf 407}, 415–419.\\\\
14. S. Mukerji and S. Chakraborty (2010), {\it Astrophys. Space
Sc.}DOI: 10.1007/s10509-010-0456-1.\\\\
15. R. Myrzakulov (2010)  arXiv:1006.1120v1 [gr-qc]\\\\
16. G. Cognola et al (2009), {\it Phys. Rev. D} {\bf 79} 044001. \\\\
17. T. Tamaki and S. Tsujikawa (2008), {\it Phys. Rev. D} {\bf 78} 084028.\\\\
18. S. Capozziello (2002), {\it Int. J. Mod. Phys. D} {\bf 11}
483.\\\\
19. S. Capozziello, V. F. Cardone, S. Carloni and A. Troisi (2003)
{\it Int. J. Mod. Phys. D} {\bf 12} 1969.\\\\
20. S. Nojiri and S. D. Odintsov (2003), {\it Phys. Rev. D} {\bf
68}123512.\\\\
21. S. Nojiri and S. D. Odintsov (2010), arXiv:1011.0544v2 [gr-qc]\\\\
22. S. Nojiri and S. D. Odintsov (2007), {\it Phys Lett B} {\bf
657} 238.\\\\
23. P. Brax et al (2008), {\it Phys Rev D} {\bf 78} 104021.\\\\
24. F. Cannata, A.Yu. Kamenshchik, arXiv:1005.1878v1 [gr-qc]
(2010).\\\\
25. K. K. Yerzhanov et al (2010), arXiv:1006.3879v1 [gr-qc].\\\\
26. R. Myrzakulov (2010), arXiv:1006.1120v1 [gr-qc].\\\\
27. E. V. Linder (2010), {\it Phys. Rev. D} {\bf 81} 127301.\\\\
28. G. R. Bengochea and R. Ferraro (2009), {\it Phys Rev D} {\bf 79} 124019.\\\\
29. S. Capozziello, V. F. Cardone and V. Salzano (2008) {\it Phys Rev D} {\it 78} 063504.\\\\
30. M. Visser (2004) {\it Class. Quantum Grav.} {\bf 21} 2603.\\\\
31. P. B. Khatua and U. Debnath (2010), {\it Astrophys. Space Sc.}
{\bf 326} 60.\\\\
32. N. Banerjee, S. Das, K. Ganguly (2010), {\it Pramana} {\bf 74} L481.\\\\
33. J. Khoury and A. Weltman (2004), {\it Phys. Rev. D} {\bf 69} 044026.\\\\
34. S. Capozziello et al (2006), {\it Phys Lett B} {\bf 639}
135.\\\\
35. S. Nojiri and S. D. Odintsov (2006), {\it Phys Rev D} {\bf 74}
086005.\\\\
36. S. Nojiri and S. D. Odintsov (2003), {\it Phys Rev D} {\bf
68}, 123512.\\\\
37. S. Nojiri and S. D. Odintsov (2004), {\it General Relativity
and Gravitation} {\bf 36} 1765.\\\\
38. G. Cognola et al (2005), {\it JCAP} {\bf 02} 10.\\\\
39. M. H. Sadjadi (2007), {\it Phys. Rev. D} {\bf 76} 104024.\\\\
40. V. Faraoni (2007), {\it Phys. Rev. D} {\bf 75} 067302.\\\\
41. S. Nojiri and S. D. Odintsov (2007), {\it Phys Lett B} {\bf 657} 238.\\\\
42. S. Nojiri and S. D. Odintsov (2008), {\it Phys Rev D} {\bf 77} 026007.\\\\
43. G. Cognola et al (2008), {\it Phys Rev D} {\bf 77} 046009.\\\\
44. S. Nojiri, S.D. Odintsov (2009), {\it AIP Conf. Proc.} {\bf
1115} 212, arXiv:0810.1557\\\\
45. K. Bamba and C-Q. Geng (2009), {\it Phys. Lett. B} {\bf 679}
282; C-J. Feng (2009), {\it Phys. Lett. B} {\bf 676}
168. \\\\
46. K. Nozari and T. Azizi (2009), {\it Phys. Lett. B} {\bf 680} 205.\\\\
47. S. K. Srivastava (2008), {\it Int J Theor Phys} {\bf 47} 1966.\\\\
48. S. Nojiri and S. D. Odintsov (2007), {\it Int J of Geometric Methods in Modern Physics} {\bf 4} 115.\\\\
49. M.C.B. Abdalla, S. Nojiri, S. D.Odintsov (2005), {\it Class. Quant. Grav.} {\bf 22} L35.\\\\
50. K. Bamba, C-Q. Geng, S. Nojiri and S. D. Odintsov (2009), {\it
Phys. Rev. D} {\bf 79}  083014.\\\\
51. K.K.Yerzhanov, Sh.R.Myrzakul, I.I.Kulnazarov, R.Myrzakulov
(2010), arXiv:1006.3879v1 [gr-qc]\\\\
52. R. Myrzakulov, arXiv:1006.1120v1 [gr-qc] (2010).\\\\
53. P. Wu, H. Yu (2010) {\it Phys Lett B} {\bf 693} 415
(2010).\\\\
54. R. Yang (2010), arXiv:1007.3571.\\\\
55. S-H. Chen, J. B. Dent, S. Dutta, E. N. Saridakis (2010),
arXiv:1008.1250v2 [astro-ph.CO].\\\\
56. J.B. Dent, S. Dutta, E.N. Saridakis (2010),
arXiv:1008.1250.\\\\
57. G. R. Bengochea (2011), {\it Phys Lett B} {\bf 695} 405.\\\\
58. G. R. Bengochea and R. Ferraro (2009), {\it Phys. Rev. D} {\bf
79} 124019.\\\\

\end{document}